\documentclass[12pt]{article}        

\usepackage{amsmath}
\usepackage{amssymb}
\usepackage{euscript}

\usepackage{epsfig}

\usepackage{cite}

\usepackage{alltt}

\usepackage{epic}

\begin{document}                     

\allowdisplaybreaks

\begin{titlepage}

\begin{flushright}
{\bf  
TPJU 10/2000 }
\end{flushright}

\vspace{2mm}
\begin{center}{\bf\Large The tauola-photos-F environment }\end{center}
\begin{center}{\bf\Large for versioning the TAUOLA and PHOTOS packages  $^{\dag}$ }\end{center}

\vspace{1mm}
\begin{center}
  {\bf   P. Golonka$^{a,b}$,}
  {\bf   E. Richter-W\c{a}s$^{a,c,d}$}
  {\em and}
  {\bf   Z. W\c{a}s$^{a,e}$ }
\\
\vspace{1mm}
{\em $^a$Institute of Nuclear Physics,
         ul. Kawiory 26a, 30-055 Cracow, Poland.}\\
{\em $^b$Faculty of Nuclear Physics and Techniques, Cracow, Poland.}\\
{\em $^c$Department of Computer Science, Jagellonian University}\\
{\em $^d$CERN, PPE, CH-1211 Geneva 23, Switzerland.}\\
{\em $^e$CERN, Theory Division, CH-1211 Geneva 23, Switzerland.}\\
\end{center}

\vspace{1mm}
\begin{abstract}
We present the system for versioning two packages: the {\tt TAUOLA} 
of $\tau$ lepton decay 
and {\tt PHOTOS} for radiative corrections in decays.
The following features can be chosen in automatic or semi-automatic way:
(1) format of the  common block {\tt HEPEVT}; (2) version of the physics
input (for {\tt TAUOLA}): as published, as initialized by CLEO collaboration, as initialized 
by ALEPH collaboration (it is suggested to use this 
    version only with the help of the collaboration advice); 
(3) type of application: stand-alone, universal interface through 
{\tt HEPEVT}, interface for {\tt KKMC} Monte Carlo; 
(4) random number generators; 
(5) compiler options.

\end{abstract}
\begin{center}
\end{center}

\vspace{1mm}
\footnoterule
\noindent
{\footnotesize
\begin{itemize}
\item[${\dag}$]
Work supported in part by Polish Government grants 
KBN 2P03B11819, 
the Maria Sk\l{}odowska-Curie Joint Fund II PAA/DOE-97-316,
and the Polish--French Collaboration within IN2P3 through LAPP Annecy.
\item[${\;}$] Home page at http://wasm.home.cern.ch/wasm/
\end{itemize}
}
\vspace{1mm}
\begin{flushleft}
{\bf 
  TPJU 10/2000  \\
  August 2000}
\end{flushleft}

\end{titlepage}

\section{Introduction}
The {\tt TAUOLA}
\cite{Jadach:1990mz,Jezabek:1991qp,Jadach:1993hs} and  
{\tt PHOTOS} \cite{Barberio:1990ms,Barberio:1994qi} are the computing
projects of rather long history. Written and maintained by the
well defined authors, nonetheless migrated into wide range
of applications where became ingredients of the
complicated simulations chains. As a consequence a large number of
different versions are presently in use.
From the algorithmic point of view, they often
differ only in a few small details, but incorporate
substantial amount of specific results from the distinct
$\tau$-lepton measurements. Often program versions differ because of the requirements of
interfaces to other packages used in the simulation chains (eg. format
of the event record has to be adjusted). 

Present utility setup for constructing specific versions of {\tt TAUOLA}
and  {\tt PHOTOS}
is prepared for the software librarians and advanced users interested in updating
both packages in the multipurpose environment. 
The idea was to create a repository which allows to include and keep main 
options of {\tt TAUOLA} developed for different purposes. 
At the same time repository can provide the standard Fortran files 
which can be handled later in
exactly the same way as the published versions of the packages.

Our present document is not aimed to be the manual of the 
{\tt PHOTOS} and {\tt TAUOLA} packages.
It is assumed that the  user is familiar with the programs themselves
and their documentation, refs. \cite{Jadach:1990mz,Jezabek:1991qp,Jadach:1993hs} and  
\cite{Barberio:1990ms,Barberio:1994qi}.
\vskip 3 mm

{\bf Motivations for versioning}:

\begin{enumerate}
\item
PHOTOS: Versions of Fortran code are necessary because of the different versions of
        the {\tt HEPEVT} common block being in use in the HEP libraries (single/double precisions,  maximal
        number of entries). 
\item
TAUOLA: Versions of Fortran code are motivated by: (A) different versions of
        initialization of physics parameters; (B)
        interfaces with different Monte Carlo generators for production of $\tau$-lepton(s); 
        and
        (C) different versions of the {\tt HEPEVT} common block:
        \begin{itemize}
        \item
(A) Different physics initializations: \\
(1) As published in \cite{Jadach:1993hs}; \\
(2) As initialized by ALEPH collaboration  \cite{aleph} (it is suggested to use this 
    version only with the help of the collaboration advice); \\
(3) As initialized by CLEO collaboration \cite{cleo} (see printout of this 
   version for details); \\
(4) Further coding of some individual decay modes.
\item
(B) Different interfaces with MC generators:\\
 (1) Old demo program as in published version \cite{Jadach:1993hs}; \\
 (2) Interface to {\tt KKMC} \cite{kkcpc:1999}; \\
 (3) New universal interface using {\tt HEPEVT}  common block.
\item
(C) Different versions of the {\tt HEPEVT} common block.\\
\end{itemize}
\item
 TAUOLA and PHOTOS: different versions of random number generators. 
\item
 TAUOLA and PHOTOS: makefiles with different compiler flags.
\end{enumerate}

The aim is to provide full backward compatibility at the level of Fortran
source with the various versions
being at present in use. Standard tools are used in the discussed setup:
cpp the C-language pre-compiler: its {\tt if}, {\tt elif} and {\tt include} commands, as well
as Unix logical links and {\tt cat} command. 
It is expected that the user will use this setup to create her/his version of {\tt TAUOLA}
and {\tt PHOTOS} libraries (subdirectories {\tt tauola/} and {\tt
photos/}) and other subdirectories of the setup will be erased/stored separately.

\section{Organization of the directory tree}

Once unpacked, the main directory  { \tt TAUOLA} is created. 

\noindent
It contains {\tt README} file and:

\vskip 2mm

\centerline{ \bf subdirectories including Fortran pre-code}

\vskip 2mm

\begin{enumerate}
\item
{\tt photos-F/}: Main directory containing PHOTOS pre-code with options.
\item
{\tt tauola-F/}: Main directory containing TAUOLA pre-code with options.
\item
{\tt demo-factory}: Directory for updating input in demo files. For specialized 
use only, see section 2.3.
\item
{\tt randg/}: Directory containing random number generators which are kept 
separately from the rest of {\tt TAUOLA} and {\tt PHOTOS} source code. 
This should facilitate replacement with the versions of random number generators 
favoured by the user. Random number generators are kept in Fortran subroutines
placed in files {\tt photos-random.h} and {\tt tauola-random.h}.
\item
{\tt include/}: Directory containing the {\tt  HEPEVT-xxx.h} files  for different versions of
the {\tt HEPEVT} common block.
The logical link {\tt HEPEVT.h} to the one actually used will be placed
in this directory later.
\end{enumerate}

\vskip 2mm

\centerline{ \bf subdirectories necessary to run demo and,} 
\centerline{ \bf to install packages on different platforms}

\vskip 2mm

\begin{enumerate}
\item
{\tt glibk/}: Directory containing histograming package, used by demos only.
\item
{\tt jetset/}: Directory containing {\tt JETSET} MC package, used by demos only.
\item
{\tt platform/}: Directory containing system-dependent versions of
                    {\tt make.inc} files for supported  platforms.
\item
{\tt make.inc}: Logical link to the chosen {\tt make-xxxx.inc} 
located in subdirectory {\tt platform/}. The {\tt make-xxxx.inc}
files define machine-dependent flags for compilers etc. to be used by all
{\tt makefiles}.
\end{enumerate}

\vskip 2mm
\centerline{\bf The following directories are created}
\centerline{\bf once the actions of the setup are completed:}
\vskip 2mm

\begin{enumerate}
  \item
  {\tt photos/}: Standard directory with Fortran code of {\tt PHOTOS}
  library and its demo.
  \item
  {\tt tauola/}: Standard directory with Fortran code of {\tt TAUOLA} library, 
                    its demos and example outputs. 
  \begin{enumerate}
    \item
    {\tt tauola/demo-standalone}: Demo program for {\tt TAUOLA} executed in a standalone mode.
    \item
    {\tt tauola/demo-jetset}: Demo program for {\tt  TAUOLA} executed with
    universal interface to physics event generators
    based on the
    {\tt HEPEVT} common block. In this demo  {\tt HEPEVT} is filled 
    from JETSET74 \cite{jetset6.3:1987} Monte Carlo generator.
   \item
   {\tt tauola/demo-KK-face}: Interface to KK Monte Carlo \cite{kkcpc:1999}. 
  \end{enumerate}
\end{enumerate}

\subsection{Options for PHOTOS Monte Carlo}
Different options of PHOTOS which can be created correspond solely to
the different versions of the {\tt HEPEVT} common block.
The possible options are:
\begin{enumerate}
\item
{\tt KK-all}     -- for KK Monte Carlo
\item
{\tt 2kD-all}    -- dimension  2000 double precision
\item
{\tt 4kD-all}    -- dimension  4000 double precision
\item
{\tt 2kR-all}    -- dimension  2000 single precision
\item
{\tt 10kD-all}   -- dimension 10000 double precision
\end{enumerate}

The action of creating required version of the library is performed  with
the help of cpp pre-compiler. It creates file
{\tt photos.f} from file  {\tt photos.F}. Once it is done,  the logical
link to  the required version of the {\tt HEPEVT} common block is created.
This link is used in construction of {\tt tauola} library, see next section.

\subsection{Options for TAUOLA Monte Carlo}

Basic options for physics initializations  are: {\tt cpc}; {\tt cleo};
{\tt aleph}. As results of the action performed by the package:
\begin{enumerate}
\item
 {\tt tauola/} subdirectory is erased;
\item
 Directory structure of  {\tt tauola/} is rebuilt; 
\begin{itemize}
\item
 {\tt tauola/} directory is filled with the Fortran code, libraries and makefiles;
\item
 {\tt tauola/demo-xx} are filled with the Fortran code of demos;
\end{itemize}
\end{enumerate}

The three possible versions of created {\tt tauola.f} correspond to
form-factors and branching ratios defined respectively as in: 
(cpc) published version of TAUOLA; (aleph) as adopted by
ALEPH collaboration, 
(cleo) as adopted by CLEO collaboration.

{\bf Remarks:}

   $\bullet$ The {\tt makefile} files are prepared to run {\tt TAUOLA}  
   within environment of
   the distribution {\tt TAUOLA} directory, however the
   templates for makefiles are compatible with these of the {\tt KK}
   Monte Carlo. Thus if {\tt tauola} directory 
   is copied into respective place  of the {\tt KKMC} distribution tree, and
   {\tt make makflag} of {\tt KK/ffbench/} is executed, it overwrites {\tt makefile} 
   file in {\tt tauola/}. The new ones are 
   produced from {\tt makefile.templ} and 
   match the {\tt KKMC} structure. 

   $\bullet$ Additional parametrizations for form-factors, which can be useful
   in some applications, are stored in the directory {\tt
   TAUOLA/tauola-F/suppl}. They are not
   {\it ready to use} and some cross checks, how they match the actual option
   of {\tt TAUOLA} library, are mandatory. At present,
   code used in refs. \cite{Abbiendi:1999cq} and \cite{Abreu:1998cn} 
   is stored there.

\subsection{How to change setting of {\tt TAUOLA} input parameters}

It is often necessary to change some of the {\tt TAUOLA}  input
parameters like branching
ratios, mass of the $\tau$-lepton, etc. It is convenient to have it done once for 
all applications i.e. {\tt demo-KK-face}, {\tt demo-jetset} and {\tt demo-standalone}.
The purpose of the {\tt demo-factory} directory is exactly that.
Here one can create the {\tt .F} files for the interfaces, by the set of paste commands
embodied in the script {\tt klej},  out of the blocks of the Fortran code. 
More precisely the following files can be recreated:
\begin{itemize}
\item
For demo-KK-face:                 {\tt ./prod/Tauface.F} 
\item
For demo-jetset:        {\tt ./prod/tauola$\_$photos$\_$ini.F} 
\item
For demo-standalone:               {\tt ./prod/taumain.F}
\end{itemize}

For details of the intialization routines, which are semi-identical in the three cases, 
see refs. 
\cite{Jadach:1990mz,Jezabek:1991qp,Jadach:1993hs}.
This requires special care from the physics point of view. In many cases
input parameters  are inter-related with the actual choice 
of form factors. The changes should be thus performed consistently.

{\bf How to proceed:}

\begin{enumerate}
\item
   Some of the routines in directory {\tt ./source}  have to be updated by hand
   first. They are stored in individual files.
   The ones which usually should not be modified are write protected.
\item
   Later execution of the script {\tt klej} will create the following files
   from the pieces stored in directory {\tt ./source} simply by pasting them 
   together:
\begin{itemize}
 \item
 {\tt  ./prod/Tauface.F}, 
\item
 {\tt  ./prod/tauola$\_$photos$\_$ini.F}, 
\item
 {\tt ./prod/taumain.F}.
\end{itemize}
  Automatic check ({\tt diff}) with the archive versions stored in
  directory {\tt ./back}  will also be executed.
\item
  Finally the following commands copy the files into appropriate places:
  \begin{enumerate}
    \item
    {\tt cp  prod/Tauface.F ../tauola-F/tauface-KK-F/Tauface.F}
    \item
    {\tt cp  prod/tauola$\_$photos$\_$ini.F ../tauola-F/jetset-F/tauola$\_$photos$\_$ini.F}
    \item
    {\tt cp  prod/taumain.F ../tauola-F/standalone-F/taumain.F}
  \end{enumerate}
\end{enumerate}

\subsection{Random number generators}

 \begin{itemize}
 \item
  {\tt PHOTOS} and {\tt TAUOLA} have their own copies of the random number generators.
  They are contained in the include files placed in the directory {\tt randg}.
 \item
  The user who wants to implement her/his own generators, eg.
 compatible with the ones used by the collaboration, should replace
 files:
 \begin{itemize}
 \item 
    {\tt ./photos-random.h},
 \item 
    {\tt ./tauola-random.h}
 \end{itemize}
  with the files including the appropriate wrappers of his own random 
  generators or empty files if the generators of the same
  name reside elsewhere.
\end{itemize} 
\subsection{Compiler flags etc}
Platform dependent parts of the {\tt makefiles} are stored
in directory {\tt platform/}.
At present options for LINUX and AIX platforms are available only. 
But it is rather straightforward to extend them to the new ones. 

\section{Universal interface with {\tt HEPEVT} common block}

Universal interface to different Monte Carlo generators is provided
through event record {\tt HEPEVT}. As a demonstration example it is
interfaced with {\tt JETSET} generator, however it should work
in the same manner with {\tt PYTHIA, HERWIG} or {\tt ISAJET} generators.
\begin{itemize}
\item
$\tau$-lepton should be forced to be stable in the event generator.
\item
Content of the {\tt HEPEVT} common block is searched for all $\tau$
leptons and neutrinos. 
\item
It is checked if there are $\tau$-flavour pairs (two $\tau$-leptons or
$\tau$-lepton and $\tau$-neutrino) originating from the same mother. 
\item
Decay of the $\tau$-flavour pairs are performed with {\tt TAUOLA}.
In some cases spin correlations are included explicitly.
This is the case of the decay of
$W$- and $Z$-bosons: $W \to \tau \nu$ and $Z \to \tau \tau$ and  decays of
scalar and pseudoscalar Higgses:
$H \to \tau \tau$, $A \to \tau \tau$, $H^{\pm}\to  \tau \nu$.
At present it is treated in the
incomplete manner only. Parallel or antiparallel spin configurations are
generated, but for $\tau \tau$ pairs originating from $Z/\gamma$ each
possibility is taken with  50 \% chances.
\item
Photon radiation in decay is performed with {\tt PHOTOS}.
\end{itemize}

Finally let us note that the calculation of the $\tau$ polarization created from
the $Z$ and/or virtual $\gamma$ (as function of the direction), represents 
rather non trivial extension.
Dedicated study of 
the production matrix elements of the host generator is necessary.
Separate paper \cite{whoever} will be devoted to this point.
\section{How to use the package}

\begin{enumerate}
\item
   Start with {\tt make Clean} from main directory 
   to secure against mismatches. 
\item
   Check platform dependent makefiles. 
\begin{itemize}
\item
    Go to subdirectory {\tt platform/}
\item
    Determine if {\tt make-xxx.inc} file specific for your computer is present there:   
    for LINUX it is {\tt make-linux.inc}; for AIX it is {\tt make-aix.inc};
    for other you need to clone/write it.
\item
    Erase symbolic link {\tt make.inc} existing in this directory and create a new
    one which points to the chosen {\tt make-xxx.inc}: 
    \begin{itemize}
     \item {\tt rm make.inc} 
     \item for LINUX: {\tt ln -s make-linux.inc make.inc} 
     \item for AIX: {\tt ln -s make-aix.inc   make.inc} 
    \end{itemize}
   Afterwards check whether link to the {\tt make.inc}
   is present in the  main directory.
\end{itemize}
\item
  Settings of {\tt TAUOLA} input parameters  can be changed  
for all implemented applications, 
see chapter 2.3 for details. 
\item
   {\tt PHOTOS} and {\tt TAUOLA} have their own private random generators. If you wish
to replace them, you should do it at this point, see chapter 2.4.
\item
Create required versions of {\tt photos/} and {\tt tauola/}  directories. It is mandatory
to create {\tt photos/} directory first, i.e. before creating {\tt tauola/}. 
\begin{itemize}
\item
    Go to directory {\tt photos-F} 
\item
    Type one of the following commands to choose the required version of {\tt HEPEVT}: 
\begin{itemize}
       \item
       {\tt make KK-all}   
       \item
       {\tt make 2kD-all}  
       \item
       {\tt make 4kD-all}  
       \item
       {\tt make 2kR-all}  
       \item
       {\tt make 10kD-all} 
\end{itemize}
\item
   Go to directory tauola-F: 
\item
   Type one of the following commands to choose the required version of {\tt TAUOLA} initialization: 
\begin{itemize}
\item
      {\tt make cpc} 
\item
      {\tt make cleo} 
\item
      {\tt make aleph} 
\end{itemize}
\end{itemize}
\item
The required version of {\tt PHOTOS} and {\tt TAUOLA} will reside in newly (re)created
directories {\tt ./photos} and {\tt ./tauola}. 
\item
Following {\tt demos} can be invoked from that directories:
\begin{itemize}
\item
   Demo for {\tt PHOTOS}  resides in {\tt ./photos/demo} 
     and can be invoked by command {\tt make} followed by {\tt make run}. 
\item
   Demo for {\tt TAUOLA}  stand-alone resides in \\
   {\tt ./tauola/demo-standalone} 
   and can be invoked by the command {\tt make} followed by  {\tt make run}.
\item
   Demo for {\tt TAUOLA} with  {\tt JETSET} being a host Monte Carlo 
   resides in \\  {\tt ./tauola/jetset-demo} and can be invoked by the 
   command {\tt make} followed by  {\tt make run}.
\item
   Interface to {\tt KKMC}  resides in 
     {\tt ./tauola/KK-face/Tauface.f}. It has to be moved to  {\tt
     ./KK2f/Tauface.f}  of  distribution directory of  KKMC
     \cite{kkcpc:1999}. The rest of the  {\tt ./tauola} directory should replace
     the original one of the 
     {\tt KK} Monte Carlo distribution.
\end{itemize}
\end{enumerate}
Finally, let us remark that most of the {\tt TAUOLA} tree is not necessary 
and can be erased at this point.
Code and makefiles of directories {\tt ./tauola} and {\tt ./photos} are 
sufficient.
To execute demo programs, either  directories {\tt ./jetset} and {\tt ./glibk} need to be kept
or replaced by the appropriate links. 
The {\tt make.inc } logical link pointing to {\tt make-xxxx.inc } file in directory 
{\tt ./platform}, 
defining appropriate compiler flags, need to be kept also.

\section{Summary and future possibilities}

We have presented the system for creating required version of {\tt PHOTOS}
and {\tt TAUOLA} packages from their master versions. The master version 
are structured  
in relatively compact form without code duplications etc.

This was the first step toward future
attempts to develop  packages without loss of their present physics content.
Some experience, collected already in that direction,
 is summarized in \cite{MsCGolonka}.
We find the question of the language translation for 
the fixed program version relatively easy. Contrary,
the question of project continuity into further upgrades
motivated by the physics, needs to be think carefully over. Matching the 
programming styles of the e.g. OO C++
experts with the strategies of testing numerical correctness of
consecutive versions is a rather crucial issue which has to be
addressed.  Tools and methods embodied 
in Fortran survive such translation with difficulty.

Necessary strategy may thus require fluency at certain moment in the 
Fortran/OO languages {\it and}
physics content of the project by the same person. Platform independent 
tools for mixing code in Fortran and OO languages might be of great help also.

\vskip 3 mm
\centerline{ \bf Acknowledgements}
\vskip 3 mm

Authors are grateful to ALEPH, CLEO, DELPHI and OPAL collaborations
for providing their appropriate versions of the whole or parts of TAUOLA 
initialization. Useful discussions and suggestions from B. Bloch, S. Jadach,
J. H. K\"uhn and 
A. Weinstein are also acknowledged. One of the authors (ZW) acknowledges
warm hospitality of the Zurich ETH L3 group at the final stage of the project
completion.
\providecommand{\href}[2]{#2}


\end{document}